# Highly-coherent stimulated phonon oscillations in a multi-core optical fiber

**(Running title: Phonon oscillations in a multi-core optical fiber)**

H. Hagai Diamandi[§], Yosef London[§], Gil Bashan, Arik Bergman, and Avi Zadok[*]

*Faculty of Engineering and Institute for Nano-Technology and Advanced Materials, Bar-Ilan University, Ramat-Gan 5290002 Israel.*

[§]These authors contributed equally.

Email addresses: hagaid@gmail.com; yoseflondon@gmail.com; gillbashan@gmail.com; bergman.arik@gmail.com; Avinoam.Zadok@biu.ac.il

[*]Corresponding author contact information: Prof. Avi Zadok. Office: +972-3-5318882; Mobile: +972-50-8520240; Fax: +972-3-7384051; Avinoam.Zadok@biu.ac.il.

**Abstract**: Opto-mechanical oscillators that generate coherent acoustic waves are drawing much interest, in both fundamental research and applications. Narrowband oscillations can be obtained through the introduction of feedback to the acoustic wave. Most previous realizations of this concept, sometimes referred to as "phonon lasers", relied on radiation pressure and moving boundary effects in micro- or nano-structured media. Demonstrations in bulk crystals required cryogenic temperatures. In this work, stimulated emission of highly-coherent acoustic waves is achieved in a commercially-available multi-core fiber, at room temperature. The fiber is connected within an opto-electronic cavity loop. Pump light in one core is driving acoustic waves via electrostriction, whereas an optical probe wave at a different physical core undergoes photo-elastic modulation by the stimulated acoustic waves. Coupling between pump and probe is based entirely on inter-core, opto-mechanical cross-phase modulation: no direct optical feedback is provided. Single-frequency mechanical oscillations at hundreds of MHz frequencies are obtained, with side-mode suppression that is better than 55 dB. A sharp threshold and rapid collapse of the linewidth above threshold are observed. The linewidths of the acoustic oscillations are on the order of 100 Hz, orders of magnitude narrower than those of the pump and probe light sources. The relative Allan's deviation of the frequency is between 0.1-1 ppm. The frequency may be switched among several values by propagating the pump or probe waves in different cores. The results may be used in sensing, metrology and microwave-photonic information processing applications.

**Introduction**

Opto-mechanical interactions in various photonic media are drawing great interest in fundamental research, as well as towards potential applications such as sensing, precision metrology, analog processing of microwave signals etc[1-5]. Light waves may stimulate mechanical (acoustic) modes of a structure through electrostriction and radiation pressure[1-5]. The mechanical waves, in turn, can modulate and scatter light via photo-elasticity and moving boundaries effects[1-5]. Coupling between optical and mechanical waves has been demonstrated in membranes, cantilevers and other nano-structures within optical cavities[6-11]; in whispering-gallery modes of micro-spheres and micro-toroids[12-24]; in photonic-integrated circuits and nano-beams in silicon[25-38] and various other crystals[39-46]; in chalcogenide glass waveguides[47-50]; in tapered fibers[51-54], micro-structured and photonic-crystal fibers[55-60]; and also in standard optical fibers. The most common form of opto-mechanical interaction in standard fiber is backwards stimulated Brillouin scattering within the core[61-63], which is widely employed in distributed sensing and microwave-photonic signal processing[64-67]. Optical fibers also support guided acoustic modes with transverse profiles that span the entire cladding cross-section[68-73]. These modes as well may interact with guided light[68-73].

One of the most striking manifestations of coupling between light and sound is stimulated oscillations of one or both wave phenomena. In Brillouin lasers[20-22,34,46,49,74-75], feedback is provided to an optical wave in a fiber, an integrated-photonic circuit or a device cavity. The resulting optical oscillations may be extremely narrowband. Some Brillouin lasers can also generate narrowband acoustic waves as a by-product[34]. In other arrangements, feedback is provided for the oscillations of the mechanical wave, instead of (or in addition to) those of the optical wave[19-20,40,76-79]. These so-called "phonon lasers" emit highly coherent acoustic waves. Most previous demonstrations relied on radiation pressure and moving boundary effects in nano-structures within an optical cavity[40], or in whispering gallery resonators[19-20]. Oscillations of optical phonons were also demonstrated in atomic media[78]. Realizations of acoustic "phonon lasers" in bulk materials required cooling to cryogenic temperatures, in order to reduce dissipation[79].

In this work, the stimulated emission of highly coherent acoustic waves in a commercially-available multi-core fiber is demonstrated at room temperature. Multi-core fibers are being developed for the parallel transport of several data channels in space-division multiplexing optical communication networks[80-81]. They are readily employed in distributed shape sensing[82], and in microwave-photonic signal processing[83-84]. Multi-core fibers also provide a rich playground for the study of opto-mechanical interactions. We have recently shown that guided acoustic waves can give rise to significant inter-core, opto-mechanical cross-phase modulation, even when the direct coupling of optical power among the cores is negligible[85]. The magnitude of the effect at specific resonance frequencies is comparable with that of intra-core cross-phase modulation due to Kerr nonlinearity[85]. Similar coupling was also demonstrated in silicon-photonic devices[32,35].

The inter-core opto-mechanical crosstalk can be driven into stimulated phonon emission in an electro-optic oscillator cavity. In these hybrid cavities, a radio-frequency waveform is modulating light at the input of an optical fiber. The waveform is detected at the fiber output, amplified and fed back to the input modulation[86-89]. Opto-electronic oscillators were recently implemented over multi-core fibers as well[90]. In nearly all opto-electronic oscillators, the fiber is only employed as a long low-loss delay line, and the propagation of light is strictly linear. The arrangement used in this work is markedly different: The radio-frequency waveform modulates pump light that is launched into the inner core of a seven-core fiber, and it is recovered upon detection of an optical probe wave that is propagating in an outer core. In the absence of opto-mechanics, no coupling may take place between pump and probe waves in separate cores, and the opto-electronic oscillator loop remains open. However, opto-mechanical inter-core crosstalk imprints the intensity modulation of the pump wave onto the probe[91,92]. The crosstalk spectrum consists of sparse, narrowband resonances[85], hence the coupling between pump and probe is highly frequency-selective. Given sufficient optical pump power, the opto-electronic cavity may reach stable, single-radio-frequency voltage oscillations[91,92].

We show below that the loop voltage is directly proportional to the material displacement magnitude of the guided acoustic waves, hence the voltage oscillations also signify similar phononic oscillations. The stimulated amplification of acoustic waves manifests in sharp threshold behavior, and in collapse of the linewidth above threshold. The linewidth is on the order of only 100 Hz, much narrower than those of the optical sources of both pump and probe. The frequency of oscillations may be switched among several values by moving the pump and/or probe waves to different cores. The proposed concept could lead to the development of practical sources of highly-coherent phonons.

# Results

## *Principle of operation*

In this section, the threshold condition for steady-state oscillations of the opto-electronic loop voltage, and the relation between the output voltage and optical pump power above threshold, are derived first. The stimulation of the guided acoustic wave in the multi-core fiber is addressed next. Finally, we show that steady-state voltage oscillations are accompanied by stimulated amplification of acoustic waves.

The principle of operation is illustrated in Fig. 1. Light from the output of a first laser diode is used as an optical pump wave. The pump wave propagates through an electro-optic amplitude modulator that is biased at 50% power transmission (quadrature). The output voltage of an opto-electronic oscillator loop is applied to the radio-frequency port of the modulator. The modulated pump wave is amplified by an erbium-doped fiber amplifier to an average optical power $P_p$. Let us denote the voltage amplitude at steady-state as $V(\Omega)$, where $\Omega$ is the radio-frequency of oscillations. The magnitude of optical pump power modulation at the fundamental frequency $\Omega$ at the optical amplifier output is given by $\delta P_p(\Omega) = J_1\left[\pi V(\Omega)/V_\pi\right]P_p$. Here $V_\pi$ represents the difference in voltage that is required to switch the modulator output between maximum and minimum transmission, and $J_1$ is the first-order Bessel function of the first kind.

The pump wave is launched into the inner core of a multi-core fiber of length $L$, in the clockwise direction only and with no feedback. During its propagation, the pump wave may stimulate radial guided acoustic modes of the fiber, denoted by $R_{0,m}$ where $m$ is an integer[68-69]. Light from a second laser diode source with optical power $P_s$ is used as an optical probe wave. The probe is launched into a Sagnac interferometer fiber loop in both directions of propagation[56,72,85]. Part of the Sagnac loop passes through an outer core of the multi-core fiber. Inter-core opto-mechanical crosstalk due to guided acoustic waves is phase-matched in the forward direction[56,68-69]. Consequently, the clockwise-propagating probe wave undergoes opto-mechanical phase modulation, whereas the counter-clockwise-propagating probe wave is unaffected[56,72]. The magnitude of the non-reciprocal phase modulation equals $\delta\varphi_m(\Omega) = \gamma_m(\Omega) L \cdot \delta P(\Omega)$, where $\gamma_m(\Omega)$ is the nonlinear opto-mechanical coefficient that is associated with mode $R_{0,m}$ and the pair of cores[85,93]:

$$\gamma_m(\Omega) = \gamma_{0,m} \frac{1}{j - 2(\Omega - \Omega_m)/\Gamma_m}. \tag{1}$$

Here $\Omega_m$ is the cut-off frequency of the guided acoustic mode $R_{0,m}$ [ref. 68-69], and $\Gamma_m$ denotes the modal linewidth. The coefficient assumes its largest magnitude on resonance $\Omega = \Omega_m$ [ref. 85,93]:

$$|\gamma_m(\Omega_m)| = \gamma_{0,m} \equiv \frac{k_0}{8n^2 c \rho_0} \frac{Q_{ES}^{(m)} Q_{PE}^{(m)}}{\Gamma_m \Omega_m}. \tag{2}$$

In Eq. (2) $\rho_0$ and $n$ are the density and refractive index of silica respectively, $c$ is the speed of light in vacuum, and $k_0$ is the vacuum wavenumber of the probe wave. $Q_{ES}^{(m)}$ denotes the spatial overlap integral between the transverse profile of the electrostrictive driving force induced by the pump wave at the inner core, and that of the material displacement in mode $R_{0,m}$ [ref. 85]. Lastly, $Q_{PE}^{(m)}$ represents the spatial overlap integral between the transverse profile of the photo-elastic perturbation to the dielectric constant due to $R_{0,m}$, and that of the optical mode of the probe wave in an outer core[85].

The non-reciprocal phase modulation of the probe wave $\delta\varphi_m(\Omega)$ is converted into intensity modulation of magnitude $\delta P_s(\Omega)$ at the output of the Sagnac loop. Polarization controllers are used to align the states of polarization of the clockwise- and counter-clockwise-propagating probe wave components at the loop output, and bias the loop for maximum intensity modulation[85,91]. The intensity of the output probe is detected by a photo-receiver of responsivity $R$ [V/W]. For the length of fiber and pump power levels used in this work, $\delta\varphi_m(\Omega) \ll 2\pi$ [ref.

85]. Under these conditions, the magnitude of the detected voltage may be approximated as $\delta V(\Omega) = R \cdot \delta P_s(\Omega) \approx \frac{1}{2} R P_s \cdot \delta \varphi_m(\Omega)$. Lastly, the detected waveform is amplified by a cascade of radio-frequency electronic amplifiers with overall voltage gain $G$. Auxiliary measurements verified that the amplifier chain is operating in the linear regime for all voltage levels that are relevant to this work, so that $G$ is independent of $\delta V$.

Steady-state oscillations require that the amplified voltage magnitude $G \cdot \delta V(\Omega)$ should return to its initial value $V(\Omega)$, leading to the following condition:

$$G \cdot \tfrac{1}{2} R P_s \cdot \gamma_m(\Omega) L \cdot J_1 \left[ \pi \frac{V(\Omega)}{V_\pi} \right] P_p = V(\Omega). \tag{3}$$

Equation (3) may be solved numerically to obtain the magnitude of steady-state voltage oscillations as a function of optical pump power. The frequency-dependent, opto-mechanical inter-core crosstalk coefficient between the pair of cores $\gamma_m(\Omega)$ may be viewed as an inline radio-frequency filter, which is part of most opto-electronic oscillator setups[86-90]. Just above threshold when $x \equiv \pi V(\Omega)/V_\pi \ll 1$, we may approximate $J_1(x) \approx \tfrac{1}{2} x$, leading to:

$$P_{p,th} = \frac{4}{\pi} \frac{V_\pi}{GRP_s \gamma_m(\Omega) L}. \tag{4}$$

Equation (4) states the pump power threshold condition $P_{p,th}$ for voltage oscillations in the opto-electronic cavity that is closed around the multi-core fiber. Oscillations do not involve any direct optical feedback of pump or probe. Let $M$ denote the modal index $m$ for which $\gamma_{0,m}$ is the largest. Oscillations would take place at the radio-frequency of a longitudinal cavity mode which is the closest to $\Omega_M$. Saturation of the output voltage occurs in accordance with the modulator transfer function $J_1(x)$.

We next discuss the implications of the steady-state voltage oscillations on the stimulation of the guided acoustic wave. Consider a radial guided acoustic mode $R_{0,m}$ at a given cross-section of the multi-core fiber. The mode is stimulated by an optical pump wave that is intensity-modulated at a radio-frequency $\Omega \approx \Omega_m$. Let us denote the slowly-varying envelope of the material displacement of modal oscillations by $A_m(t)$, where $t$ stands for time. The complex envelope is governed by the following differential equation[68,69,85,94]:

$$\frac{dA_m(t)}{dt} = -\frac{\Gamma_m}{2} A_m(t) + \frac{Q_{ES}^{(m)}}{j8nc\rho_0\Omega_m} \delta P_p. \tag{5}$$

The first term on the right-hand side represents acoustic losses, which are dominated by dissipation to the fiber coating[72,73]. The second term represents an electrostrictive driving force[68,69,85]. $\delta P_p$ denotes the magnitude of the instantaneous power modulation of the pump wave, and $|\Omega - \Omega_m| \ll \Gamma_m$ is assumed. Subject to steady-state voltage oscillations of the opto-electronic loop, we may relate $\delta P_p$ with the magnitude of opto-mechanical phase modulation of the probe wave using Eq. (3):

$$\delta P_p = J_1\left(\pi \frac{V}{V_\pi}\right) P_p = \frac{V}{G \cdot \frac{1}{2} R P_s \cdot \gamma_m(\Omega) L} = \frac{\delta\varphi_m}{\gamma_m(\Omega) L} \approx j\frac{\delta\varphi_m}{\gamma_{0,m} L}. \tag{6}$$

The phase modulation, in turn, is proportional to the displacement magnitude[85]:

$$\delta\varphi_m = \frac{A_m}{2n} Q_{PE}^{(m)} k_0 L. \tag{7}$$

Equations (6) and (7) signify that steady-state operation of the opto-electronic oscillator is also associated with steady-state oscillations of the guided acoustic wave in the transverse fiber cross-section. The loop voltage magnitude is proportional to that of the acoustic displacement:

$$V = \frac{GRP_s k_0 L Q_{PE}^{(m)}}{4n} A_m. \tag{8}$$

Hence measurements of the loop output voltage directly provide information regarding the magnitude and spectrum of stimulated guided acoustic waves.

The above relation between voltage and acoustic displacement can be viewed in additional perspective. Substituting Eq. (7) into Eq. (6) yields:

$$\delta P_p = \frac{j}{\gamma_{0,m}} \frac{k_0 Q_{PE}^{(m)}}{2n} A_m, \tag{9}$$

leading to the following form of Eq. (5):

$$\frac{dA_m}{dt} = -\frac{\Gamma_m}{2} A_m + \frac{Q_{ES}^{(m)}}{8nc\rho_0 \Omega_m} \frac{k_0 Q_{PE}^{(m)}}{2n\gamma_{0,m}} A_m. \tag{10}$$

Equation (9) shows that at steady-state operation of the opto-electronic oscillator loop, the electrostrictive driving force becomes proportional to the material displacement magnitude. That term therefore represents *stimulated amplification* of the acoustic wave:

$$\frac{dA_m}{dt} = -\frac{\Gamma_m}{2} A_m + \frac{g_m}{2} A_m, \tag{11}$$

with an effective gain coefficient:

$$g_m \equiv \frac{k_0 Q_{ES}^{(m)} Q_{PE}^{(m)}}{8n^2 c\rho_0 \Omega_m} \frac{1}{\gamma_{0,m}} = \Gamma_m. \tag{12}$$

The last equality in Eq. (12) relies on Eq. (2). The equivalent gain coefficient of the acoustic waves above the threshold of voltage oscillations is locked to a constant value, which offsets losses exactly and does not depend on the optical pump power or loop voltage. This property is characteristic of laser systems. The feedback amplification of the acoustic waves is in contrast to the single-pass propagation of both pump and signal optical waves. The analysis therefore suggests that: a) sufficient optical pump power can drive the opto-electronic loop that goes through the multi-core fiber into radio-frequency voltage oscillations; and b) such oscillations go hand-in-hand with stimulated amplification of guided acoustic waves in the fiber transverse cross-sections.

*Experimental results*

Figure 2 shows examples of the radio-frequency spectrum of the loop output voltage. When the average power of the optical pump wave was below a threshold value of 2.5 W, the voltage spectra consisted of broad peaks that correspond to longitudinal modes of the opto-electronic loop. The peaks were separated by a free spectral range of about 2 MHz, and their widths were several hundreds of kHz. The three peaks shown in the figure were of equal magnitudes. When the pump power exceeded 2.5 W, narrowband oscillations appeared at a frequency of 369.4 MHz. This value matches the cut-off frequency $\Omega_8$ of radial mode $R_{0,8}$ [ref. 68, 85]. Previous work had shown that this mode introduces the strongest opto-mechanical crosstalk between the pair of cores[85]. The frequency of oscillations therefore agrees with expectations.

The peak power of the output voltage spectrum increased by more than 50 dB when the optical pump power was raised by just 1 dB above threshold. Competing longitudinal modes of the oscillator loop were suppressed by 60 dB. Spectral peaks near the cut-off frequencies $\Omega_m$ of competing guided acoustic modes were lower than the main peak by 55 dB or more (see Fig. 3). A peak near the second harmonic $2\Omega_8$ was 40 dB weaker than the main peak (Fig. 3), and third-order or higher harmonics could not be observed. The measured peak power of the voltage spectrum is plotted as a function of the optical pump power in Fig. 4, alongside corresponding calculations of Eq. (3). The parameters used in the calculations are specified in the Methods section. A sharp threshold is observed at an optical pump power of 2.5 W, as noted above. The peak power saturates at an optical pump power of 3 W. Good agreement is observed between model and experiment.

Figure 5 shows a magnified view of the normalized output voltage spectrum, in the vicinity of the 369.4 MHz peak of Fig. 2 (blue trace). The average pump power was 3.5 W. The full-width of the spectrum 20 dB below the peak was only 300 Hz. This linewidth is narrower than those of the laser diodes used as sources of pump and probe waves. The full-width at half-maximum could not be measured directly with sufficient certainty due to variations of the instantaneous frequency of oscillations (see discussion of Allan's deviation below). A Lorentzian fitting to the measured spectrum suggests a full-width at half-maximum of only 60 Hz. The phase noise of the oscillations was measured as -71 dBc/Hz at a frequency offset of 10 kHz.

To further illustrate the role of coherent acoustic oscillations, the probe wave source was replaced with a different laser diode of 1543 nm wavelength and a linewidth of 1 MHz. The normalized radio-frequency spectrum of the loop voltage in the vicinity of the main peak is shown in the red trace of Fig. 5. The pump power was again 3.5 W. The spectrum is nearly indistinguishable from the one obtained using the much narrower probe wave source (blue trace). The voltage (and acoustic displacement) waveforms are not replicas of the incident probe wave: their linewidth in this case is about 4 orders of magnitude narrower. The spectrum represents the beat note between the incident probe wave laser light, and spectral side-lobes that are generated by the acoustic wave modulation[34]. The measured narrow linewidth indicates that the spectral side-lobes are nearly-identical copies of the incident probe wave[34]. These sidebands are therefore generated by acoustic waves that are extremely coherent.

The coherence of the acoustic oscillations does not replicate that of the pump wave either. To demonstrate, the pump laser diode was replaced by a broadband amplified spontaneous emission source of 4 nm bandwidth (about 500 GHz). The probe wave laser was the 1 MHz-wide laser diode used in the previous acquisition. The normalized power spectrum of the loop voltage is shown in the black trace of Fig. 5. Narrowband oscillations were still observed, even though the pump light was extremely incoherent. The linewidth was only slightly broader than those obtained with a narrowband pump laser diode.

Allan's deviation of the instantaneous frequency of oscillations was measured using a frequency counter[96-97], on time scales between several seconds and two hours (Fig. 6). The relative frequency deviation varied between 30-200 Hz, or 0.1-1 ppm of the mean frequency. The variations represent the inherent residual instability of the oscillator setup. It should be noted that changes in the laboratory temperature modify the frequencies of longitudinal modes of the opto-electronic oscillator cavity by 7.5 ppm per °K [ref. 98]. The frequency deviations are therefore comparable with those that are induced by temperature fluctuations of less than 0.1 °K. The temperature control of the laboratory environment cannot maintain this level of stability over minutes and hours. It is therefore possible that the measurements of Allan's deviation have been affected by temperature changes, and that the inherent stability of the instantaneous frequency of oscillations might have been higher.

Lastly, the pump wave was moved from the inner core of the multi-core fiber to the 12 o'clock outer core (see legends of Fig. 7), and the probe wave was moved among three different outer cores (see Fig. 7). Each choice of core resulted in a different frequency of oscillations: 391.6 MHz, 421.3 MHz and 526.6 MHz for probe waves at the 6 o'clock, 8 o'clock core and 10 o'clock outer cores, respectively (panels 7(a) through 7(c)). The frequencies of oscillations match pronounced spectral peaks of inter-core opto-mechanical crosstalk between the pairs of cores, which were measured in previous work[85,95]. In all cases, narrowband oscillations were observed, with strong suppression of competing acoustic and longitudinal modes. The simultaneous launch of pump waves in multiple cores might lead to stimulated oscillations of additional acoustic modes.

The spectra of opto-mechanical cross-phase modulation between any pair of outer cores consist of contributions of hundreds of torsional-radial guided acoustic modes of the fiber[85]. The obtained spectra are quasi-continuous and irregular, and contributions of individual modes are difficult to pin-point. That being said, the frequency of oscillations obtained with the probe wave at the 10 o'clock outer core may be associated with the guided torsional-radial acoustic mode $TR_{18,14}$ [ref. 85,95]. The calculated transverse profile of photo-elastic perturbations to the dielectric constant due to that mode is illustrated in Fig. 7(d).

**Discussion**

The observed narrowband voltage oscillations are direct indication of equally-coherent stimulated guided acoustic waves in transverse cross-sections of the fiber. This conclusion is supported by several arguments: a) Analysis shows that the loop voltage at steady-state oscillations is simply proportional to the magnitude of acoustic displacement; b) The electrostrictive driving force of the acoustic wave stimulation at the steady-state becomes a displacement gain term, which is locked to a fixed value and offsets acoustic losses exactly; c) The frequency of oscillations matches that of the most pronounced inter-core opto-mechanical crosstalk; d) The optical paths of both pump and probe waves do not include any feedback; e) The oscillations coherence does not replicate that of the probe wave: the observed linewidth can be four orders of magnitude narrower than that of the probe laser source; and f) The oscillations coherence does not stem from that of the pump wave either. The pump wave could be an amplified spontaneous emission source with a linewidth that was nine orders of magnitude broader than that of the observed oscillations.

The principle of phononic oscillations based on opto-mechanical crosstalk between optical waveguides is equally applicable to micro-cavity and integrated-photonic devices platforms. For example, Lee *et al.* reported electrical actuation of an opto-mechanical micro-toroid in a feedback loop[99], and Shin and coworkers showed photon-phonon emit-receive operation between a pair of optical waveguides within a common membrane structure[35]. The use of integrated-photonic platforms would provide stronger opto-mechanical coupling, higher frequencies of oscillations, greater design flexibility and smaller form factors. On the other hand, the phase noise of the oscillator in a short feedback loop is expected to be higher. Large design freedom, long interaction lengths and high frequencies of operation may also be achieved in photonic crystal fibers containing multiple cores[100].

The phase noise performance of the oscillator reported in this work was rather modest: -71 dBc/Hz at a frequency offset of 10 kHz. The phase noise is orders-of-magnitude higher than those of state-of-the-art opto-electronic oscillators[101]. It should be noted, however, that the primary objective of this study has been the stimulated amplification of sound waves. Standard opto-electronic oscillators do not involve guided acoustic waves. A single-mode fiber delay line may be added as part of the opto-electronic cavity, between the output of the Sagnac loop and the photo-receiver. The low-loss delay line would reduce the losses of the oscillating electrical waveform per unit time, leading to narrower linewidth with lower phase noise[86,87]. On the other hand, a longer opto-electronic cavity would reduce the free spectral range between longitudinal modes, and might degrade the side-mode suppression. The oscillations linewidth and phase noise may also be reduced if the multi-core fiber is kept at low temperature[102]. The threshold power of 2.5 W can be lowered using longer multi-core fibers. A longer multi-core fiber can also be leveraged towards a smaller voltage gain within the opto-electronic cavity. Weaker electronic gain, in turn, may reduce the phase noise of the oscillator as well. The opto-mechanical inter-core crosstalk can be made stronger if the acrylate coating is taken off the fiber[72].

Does the acoustic oscillator constitute a "phonon laser"? The answer would largely depend on how one wishes to define that term. Previous claims for "phonon lasing" typically described arrangements in which feedback is provided to the acoustic oscillations strictly within a mechanical cavity[19-20,40,79]. In this work the acoustic wave stimulation is aided by electrical gain in an opto-electronic loop outside the fiber cavity[99]. On the other hand, the setup shares the most significant attributes of device-level "phonon lasers": it generates extremely coherent, stimulated acoustic oscillations, and it is pumped by light without optical feedback. The degree of coherence that is achieved surpasses those of "phonon laser" devices. The fiber-based acoustic oscillator may be integrated within a fiber-optic distribution network for the collection and transmission of data.

The proposed concept may lead to practical sources of highly-coherent acoustic waves at sub-GHz and even few-GHz frequencies. On top of the interest of basic research, such sources might be used for precision metrology, acoustic spectroscopy, sensing, and microwave-photonic signal processing. The acoustic waves can propagate through commercially-available coating layers, such as polyimide[103], and probe the surroundings of the fiber. The frequencies of oscillations may be affected by the properties of outside media[72]. Future work would look to explore these possibilities.

**Materials and methods**

The detailed experimental realization is shown in Fig. 8. The pump light source was a laser diode at 1559 nm wavelength, with a specified linewidth of 100 kHz. The $V_\pi$ value of the pump wave amplitude modulator was 3.5 V. The average power $P_p$ of the optical pump wave at the fiber amplifier output was varied between 1-3.5 W. The probe wave was drawn from a second laser diode at 1550 nm wavelength and 1 kHz linewidth. The average optical power of the probe wave $P_s$ at the Sagnac loop input was 3.7 mW. The overall responsivity $R$ of the photo-detector used at the Sagnac loop output and its internal circuitry was 27 V/W. The overall voltage gain $G$ of the radio-frequency amplifier chain was 1,200. Voltage test-points 1, 2 and 3 were located at the output of the photo-detector, in between two radio-frequency electronic amplifiers, and at the modulator input, respectively (see Fig. 8). The output voltage was characterized by an analysis unit, which consisted of a sampling oscilloscope, a radio-frequency electrical spectrum analyzer and a frequency counter.

The length of the multi-core fiber $L$ was 30 m. The fiber consisted of an inner core and six outer cores. The centers of outer cores were located 35 µm away from the fiber axis, on a hexagonal grid (see illustration in Fig. 1). The mode field diameter in all cores was specified as 6.4 µm. The outer diameter of the cladding was 125 µm. The fiber was coated with standard, dual-layer acrylate coating. The geometrical parameters of the fiber were used in calculations of $Q_{ES}^{(m)}$, $Q_{PE}^{(m)}$ and $\gamma_{0,m}$ [ref. 85]. The spectrum of opto-mechanical cross-phase modulation between the inner core and outer cores of the fiber was characterized in a previous study[85,95]. Analysis and measurements showed that the largest opto-mechanical inter-core cross-phase modulation is obtained through radial mode $R_{0,8}$, at a resonance frequency $\Omega_8$ of $2\pi \cdot 369.2$ MHz[85]. The magnitude of the inter-core nonlinear coefficient $\gamma_{0,8}$ at that frequency is on the order of 0.7 [W×km]$^{-1}$ [ref. 85]. The modal linewidth $\Gamma_8$ was measured as $2\pi \cdot 5.5$ MHz[72,85]. End-to-end insertion losses of optical power through the multi-core fiber and input/output fan-out units were 3-4 dB, depending on the choice of core. The residual coupling of optical power between any pair of cores, in the fiber itself and the fan-out units combined, was measured as -40 dB or lower. Optical bandpass filters were used within the Sagnac loop to further suppress the leakage of pump light towards the photo-receiver.

**Data Availability:**

All data generated or analyzed during this study are included in this article.

**Acknowledgements:**

This work was supported in part by a Starter Grant from the European Research Council (ERC): grant number H2020-ERC-2015-STG 679228 (L-SID), and by the Israeli Ministry of Science and Technology: grant number 61047.

**Author contributions:**

HHD, YL, GB, AB and AZ developed the concept and theory. HHD and YL carried out the analysis and numerical simulations. HHD, YL and GB performed measurements and analyzed data. AZ proposed the idea, managed the project and wrote the manuscript.

**Conflict of interests:**

The authors declare no competing interests.

**Figures and Figure Legends**

Figure 1:

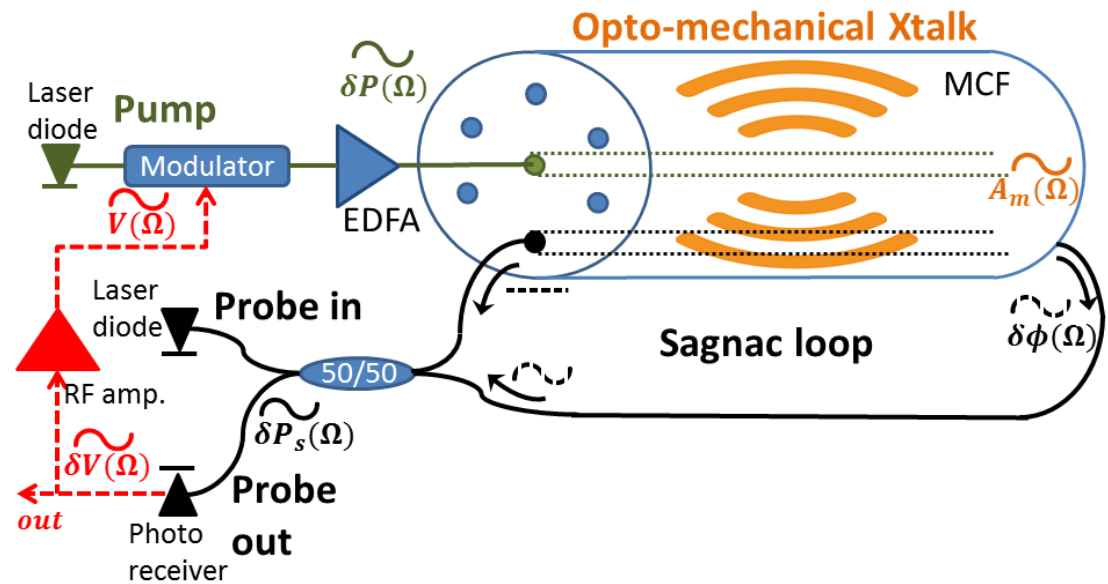

Fig. 1. Schematic illustration of the principle of operation of a stimulated acoustic wave oscillator in a multi-core fiber, within an opto-electronic loop. Pump light from a laser diode is amplitude-modulated by the radio-frequency voltage of the oscillator loop. The pump wave is amplified and launched into the inner core of a multi-core fiber. The propagation of the pump light stimulates radial guided acoustic modes of the fiber. A signal wave from a second laser diode propagates in an outer core of the fiber in both directions. The signal component that is co-propagating with the pump wave is phase-modulated by guided acoustic waves, whereas the counter-propagating signal wave is unaffected. The non-reciprocal, opto-mechanical phase modulation of the probe wave is converted to an intensity waveform at the output of a Sagnac loop. The intensity waveform is detected, amplified and fed back to modulate the optical pump wave. The opto-mechanical, inter-core crosstalk couples between the pump and signal optical waves at specific resonance radio-frequencies. Sufficient pump power can drive the loop voltage into stable, single-frequency oscillations. The measured voltage is directly proportional to the displacement magnitude of the stimulated acoustic waves. The definitions of symbols are the following: $\Omega$ - radio-frequency of oscillations; $\delta P_p(\Omega)$ - magnitude of pump power modulation; $A_m(\Omega)$ - magnitude of material displacement of the oscillations of radial guided acoustic mode $R_{0,m}$; $\delta \varphi_m(\Omega)$ - magnitude of the opto-mechanical phase modulation of the probe wave; $\delta P_s(\Omega)$ - magnitude of the probe power modulation at the output of the Sagnac loop; $\delta V(\Omega)$ - voltage magnitude at the output of the photo-receiver; $V(\Omega)$ - amplified, output loop voltage.

Figure 2.

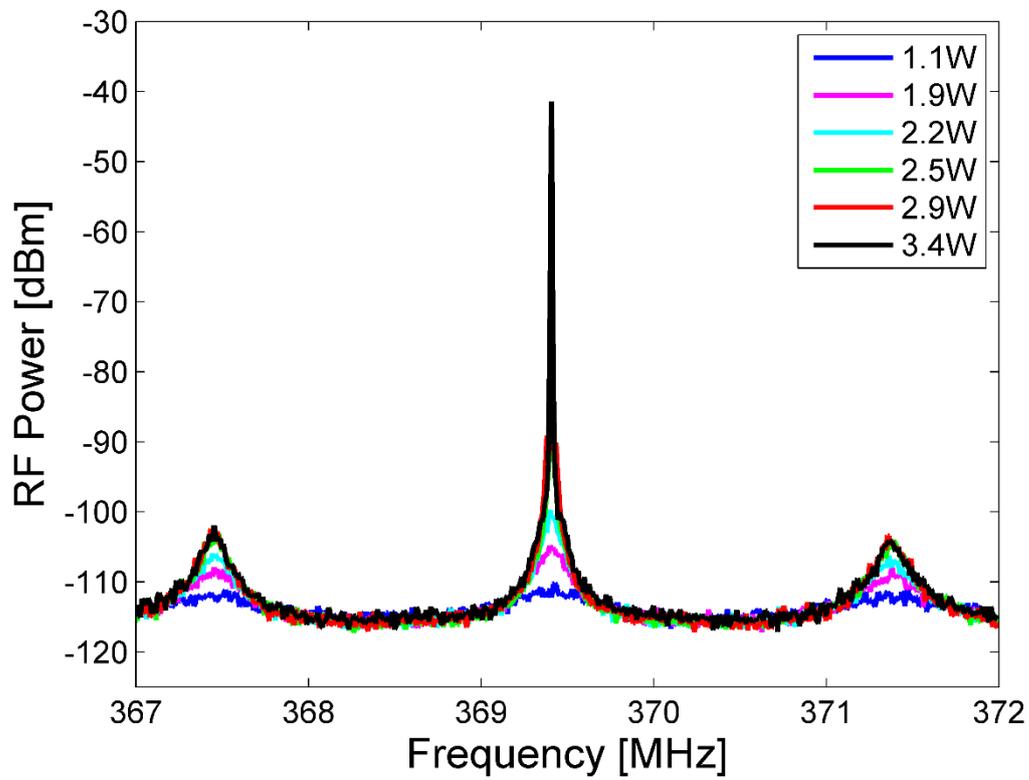

Fig. 2. Radio-frequency (RF) power spectra of the opto-electronic loop voltage, at several optical pump power levels (see legend). Oscillations at a frequency of 369.4 MHz are obtained above a threshold pump power of 2.5 W.

Figure 3.

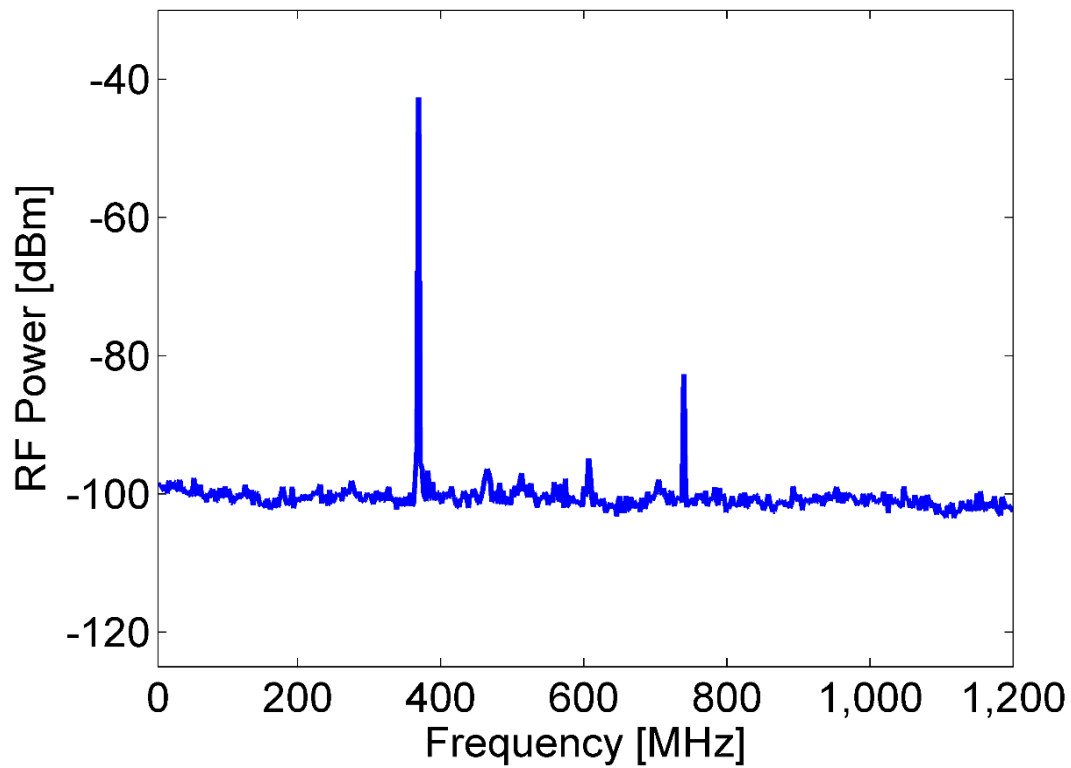

Fig. 3. Radio-frequency (RF) power spectrum of the opto-electronic loop voltage, shown on a wider frequency scale than that of Fig. 2 above. The pump power was 3 W. A main peak at the cut-off frequency $\Omega_8/(2\pi)$ = 369 MHz of guided radial acoustic mode $R_{0,8}$ is observed. Spectral peaks corresponding to competing guided acoustic modes are suppressed by at least 55 dB. A peak at the second-harmonic of $\Omega_8/(2\pi)$ is 40 dB lower than the main peak.

Figure 4.

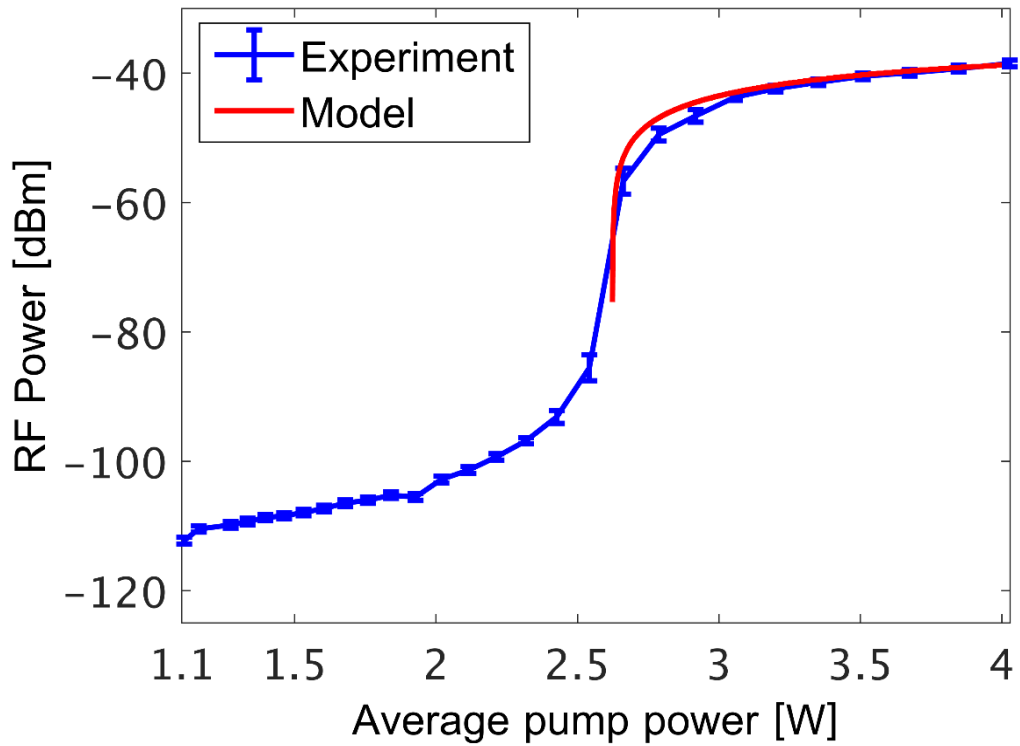

Fig. 4. Blue markers and trace: measured peak power of the voltage spectrum at the output of the opto-electronic oscillator loop, as a function of the optical pump power. A threshold is observed at a pump power of 2.5 W, followed by rapid increase in the spectral peak power by 5 orders of magnitude, and saturation at an optical pump power of about 3 W. Red trace: calculated peak power of the voltage spectrum as a function of the optical pump power above threshold (see Eq. (3)). Good agreement is obtained between model and measurement.

Figure 5.

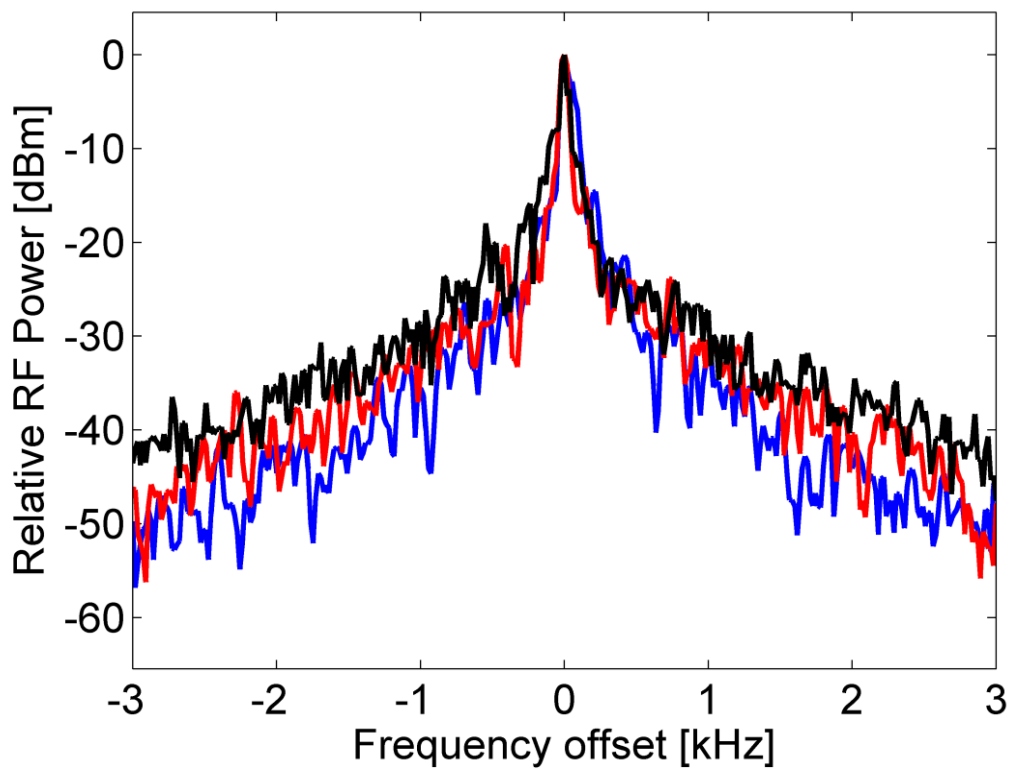

Fig. 5. Magnified view of the measured, normalized radio-frequency (RF) power spectra of the loop output voltage. The horizontal axis denotes the frequency offset from the spectral peak at 369.4 MHz (see Fig. 2). Blue trace: probe wave taken from a laser diode with a linewidth of 1 kHz. Pump wave drawn from a laser diode with 100 kHz linewidth. Red trace: probe wave source replaced by a different laser diode, with a linewidth of 1 MHz. The full widths of both spectra 20 dB below the peak are practically identical, on the order of 300 Hz. Black trace: pump laser replaced with an amplified spontaneous emission source of 500 GHz (4 nm) bandwidth. Narrowband oscillations are still obtained, with a linewidth that is only marginally broader. The optical pump power was 3.5 W in all measurements.

Figure 6.

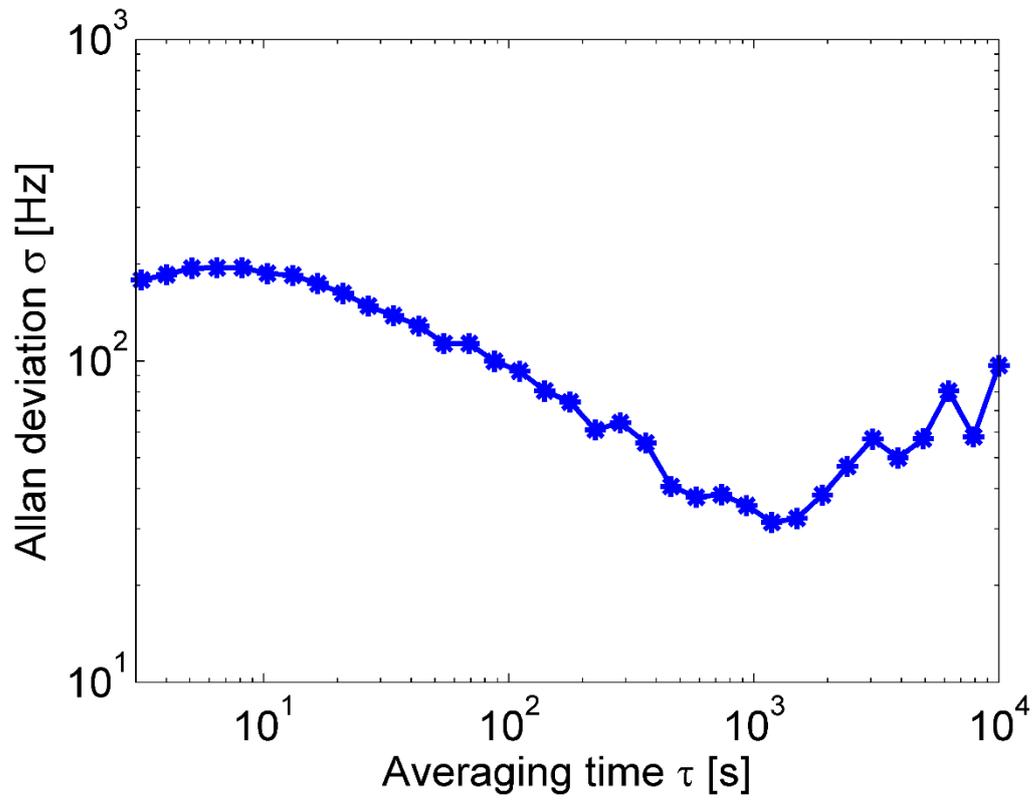

Figure 6. Measured Allan's deviation of the instantaneous frequency of voltage oscillations, as a function of the duration of integration window.

Figure 7.

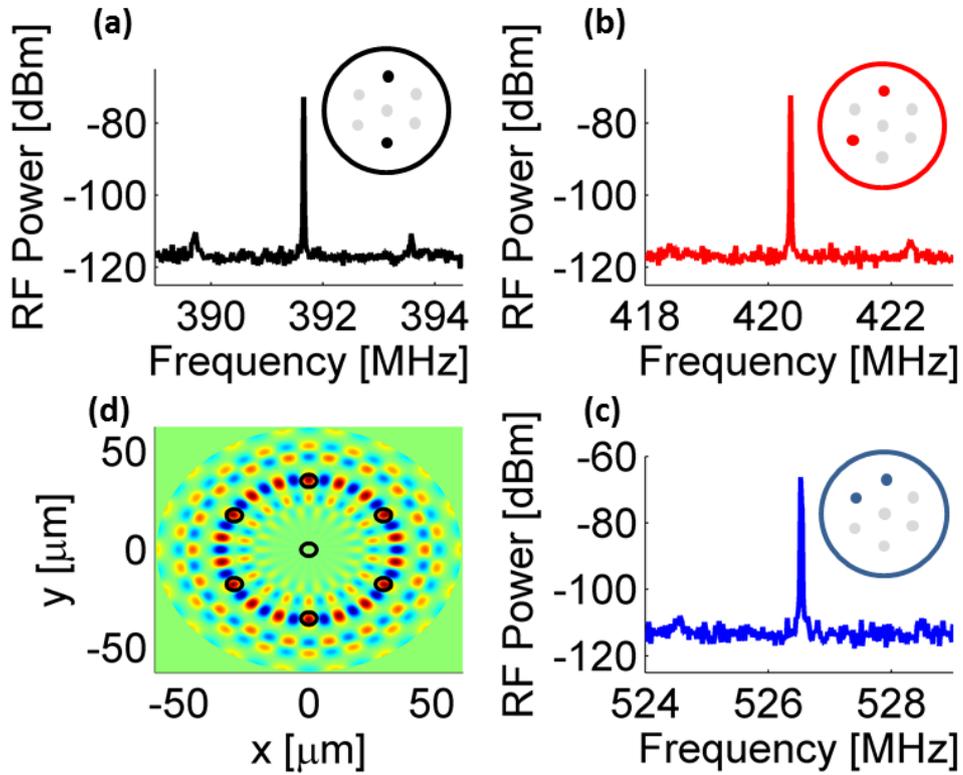

Fig. 7. (a) – Measured radio-frequency (RF) power spectrum of the opto-electronic oscillator loop voltage. The pump was propagated at the 12 o'clock outer core (see inset), and the probe was propagated in the opposite, 6 o'clock outer core. (b) – Same as panel (a), with the probe moved to the 8 o'clock outer core. (c) – Same as panel (a), with the probe moved to the 10 o'clock outer core. (d) – Calculated transverse profile of one element out of the tensor of photo-elastic perturbations to the dielectric constant of the fiber, which are induced by the guided torsional-radial acoustic mode $TR_{18,14}$ [ref. 85,95]. Peaks of photo-elastic perturbations are in spatial overlap with the outer cores. The cut-off frequency of that mode is 526 MHz, in agreement with the frequency of oscillations observed in panel (c).

Figure 8.

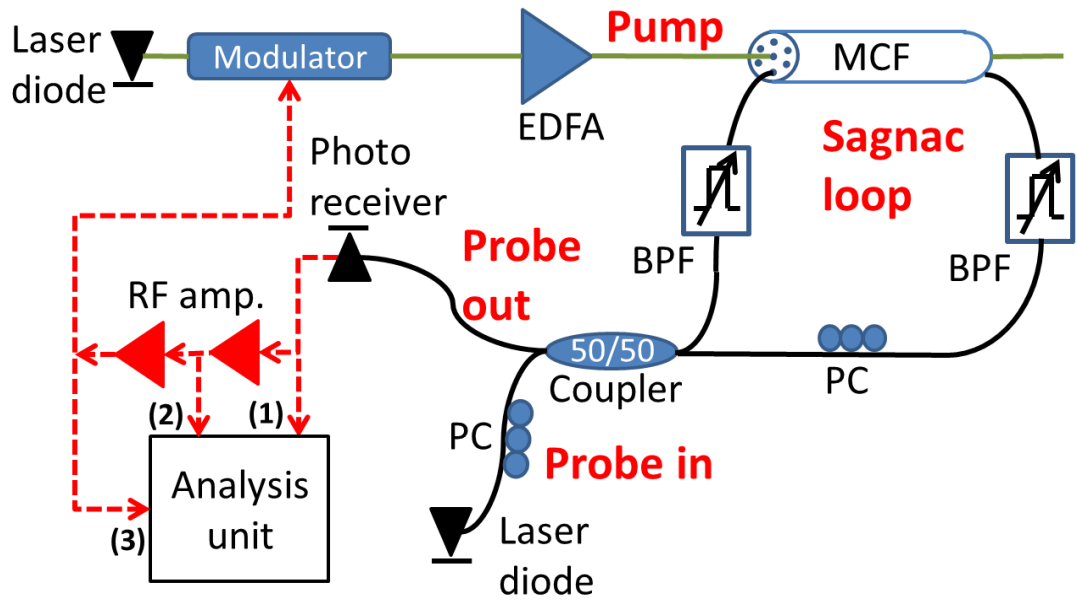

Fig. 8. Schematic illustration of the experimental setup. MCF: multi-core fiber. PC: polarization controller. BPF: tunable optical bandpass filter. RF amp.: radio-frequency voltage amplifier. EDFA: erbium-doped fiber amplifier. Voltage test-points are numbered (1), (2) and (3). Solid, green lines: fiber paths of optical pump wave. Black, solid lines: fiber paths of optical signal wave. Red, dashed lines: radio-frequency cables.